\documentclass[aps]{revtex4}
\usepackage{graphicx}
\begin{document}

\title{An On Line Interactive Geometric Database: \\ Including Exact
Solutions of Einstein's Field Equations}
\author{Mustapha Ishak\cite{email} and Kayll Lake\cite{email2}}
\affiliation{Department of Physics, Queen's University, Kingston,
Ontario, Canada, K7L 3N6 }
\date{\today}

\begin{abstract}
We describe a new interactive database (GRDB) of geometric objects
in the general area of differential geometry. Database objects
include, but are not restricted to, exact solutions of Einstein's
field equations. GRDB is designed for researchers (and teachers)
in applied mathematics, physics and related fields. The flexible
search environment allows the database to be useful over a wide
spectrum of interests, for example, from practical considerations
of neutron star models in astrophysics to abstract space-time
classification schemes. The database is built using a modular and
object-oriented design and uses several Java technologies (e.g.
Applets, Servlets, JDBC).  These are platform-independent and well
adapted for applications developed to run over the World Wide Web.
GRDB is accompanied by a virtual calculator (GRTensorJ), a
graphical user interface to the computer algebra system GRTensorII
used to perform on line coordinate, tetrad or basis calculations.
The highly interactive nature of GRDB allows for systematic
internal self-checking and a minimization of the required internal
records. This new database is now on line at
\textit{http://grdb.org}.

\end{abstract}
\maketitle
\section{Introduction}

The study of exact solutions of Einstein's Field Equations is an
important part of the theory of General Relativity \cite{kramer}
\cite{krasin}. This importance derives not only from more formal
mathematical aspects associated with the theory ( e.g. the
classification of space-times) but also from the growing
importance of the application of general relativity to
astrophysical phenomena. For example, exact solutions can offer
physical insights that numerical solutions can not. There are of
course a very large number of exact solutions offered in the
literature. However, these include numerous re-discoveries,
confusions and errors. In a recent study of static spherically
symmetric perfect fluid solutions, for example, only 9 of 127
candidate solutions were found to be of physical interest
\cite{delgaty}.

It is not surprising that some of the earliest applications of
computer algebra can be found in General Relativity
\cite{D'Inverno80} \cite{MacCallum96}. Indeed, no effective study
of exact solutions in the large can be carried on without it. With
the recent rapid development of the World Wide Web (WWW) as an
means of information transfer, it has become an effective medium
for cataloguing exact solutions. The first such use of the WWW is
due to J. Skea \cite{skea}. Skea's database is primarily a
classification database \cite{classification}. It is static in the
sense that the information that can be retrieved is resident in
the records.

In this paper we report on a non-static WWW based database, one
for which the information that can be retrieved is not restricted
to information resident in the records. This development has
become feasible using more recent WWW technologies. This new
database is now on line at \textit{http://grdb.org} (henceforth
GRDB) using several Java technologies and following a fully
modular object-oriented design. The dynamic nature of the database
is accomplished via the inclusion of GRTensorJ \cite{grtj}. This
is an interactive programmable graphical user interface to the
computer algebra system GRTensorII \cite{grtensor} which in turn
runs under Maple \cite{maple}.

GRDB is designed to be a database of geometric objects in the
general area of differential geometry with no a priori restriction
on the number of dimensions. It contains exact solutions of
Einstein's field equations without being restricted to them. By
design the database is highly interactive. The records (manifolds)
are generally stored only in terms of a metric or a set of basis
vectors along with constraints and referencing data. Other
elements can be calculated and displayed as required either
interactively or from an extensive menu.

The database is designed for researchers (and teachers) in applied
mathematics, physics and related fields and it is hoped that the
database will find use over a wide spectrum of interests.

\section{Description and Usage}

\begin{center}
\begin{figure}
\includegraphics{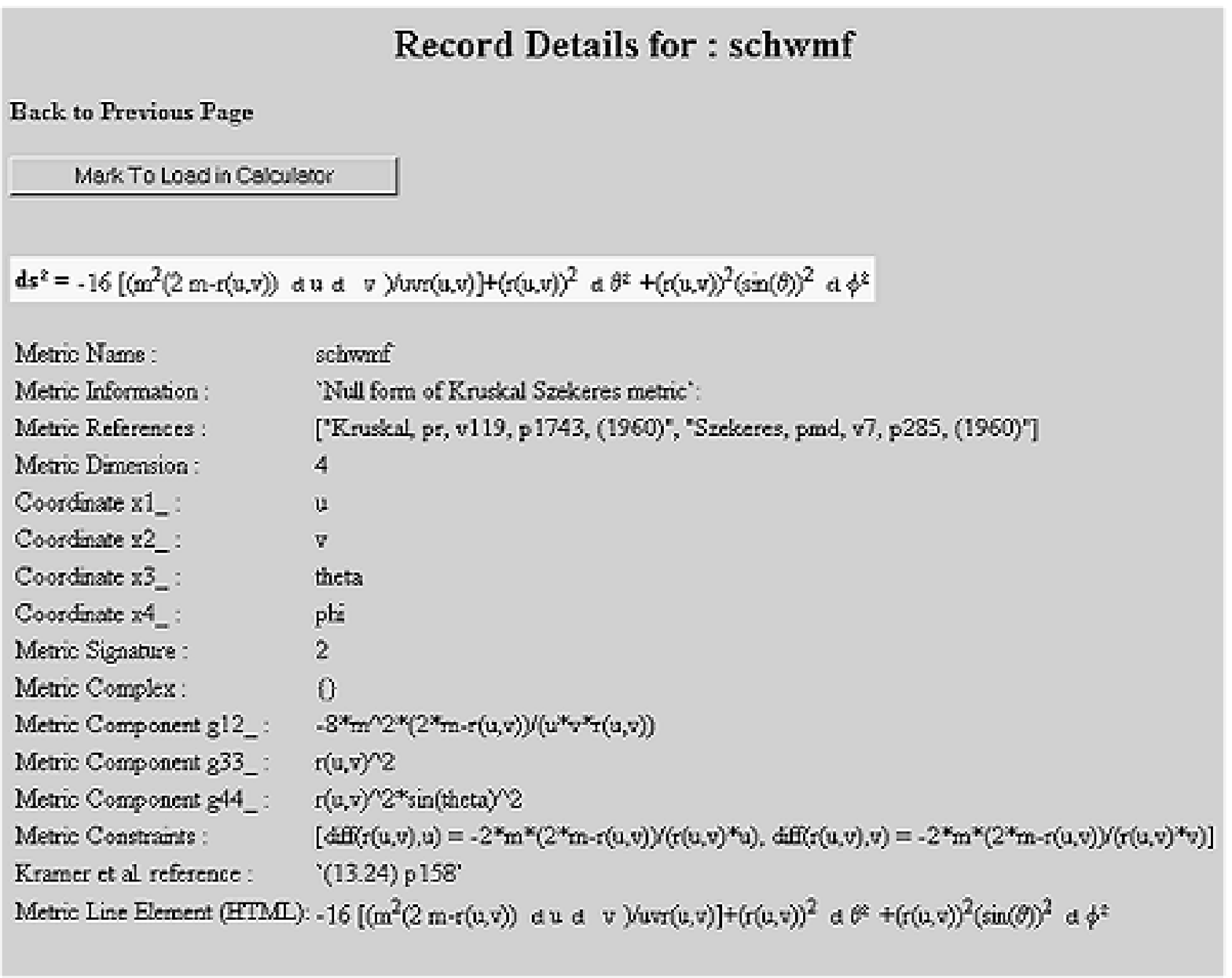}
\vspace{8pt} \caption{\label{recordpage} Details Of The Manifold's
Record. The corresponding line element is displayed in a suitable
``Tex to Html''  mathematical format. The fields contain the
components of the metric tensor, the referencing data and an
information line on the manifold. The button at the top serves to
mark the record to be loaded in the Calculator.}
\end{figure}
\end{center}

\begin{center}
\begin{figure}
\includegraphics{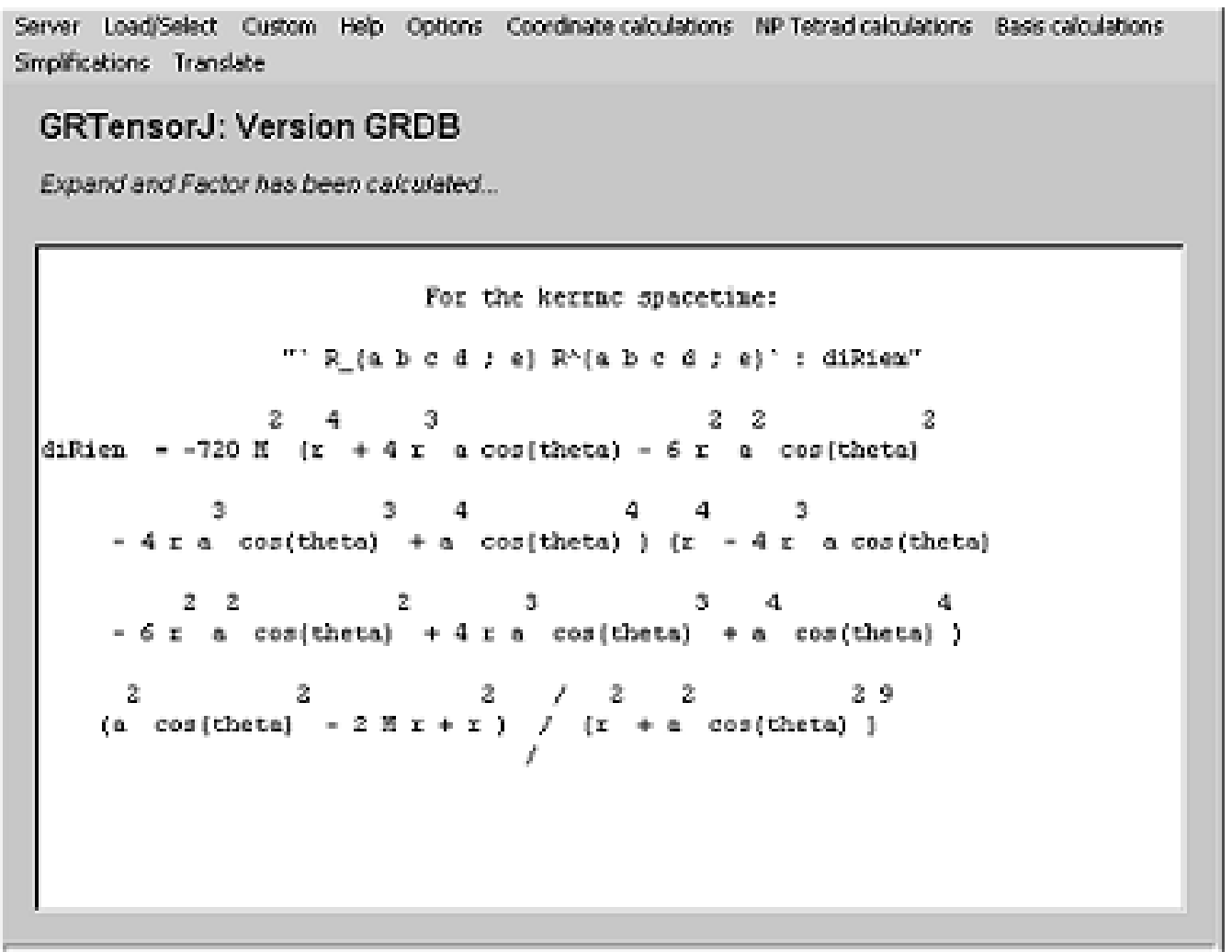}
\vspace{8pt} \caption{\label{calculator} GRTensorJ Graphical User
Interface. Calculation of the differential invariant (``diRiem'')
$R_{a b c d ; e}R^{a b c d ; e}$ for the Kerr metric is shown. At
the time of writing, this calculation executes in under one second
on a contemporary PC. The calculation commands are selected from
menus and sub-menus. The help is embedded in the menu and appears
under ``Explanation''.}
\end{figure}
\end{center}

\begin{center}
\begin{figure}
\includegraphics{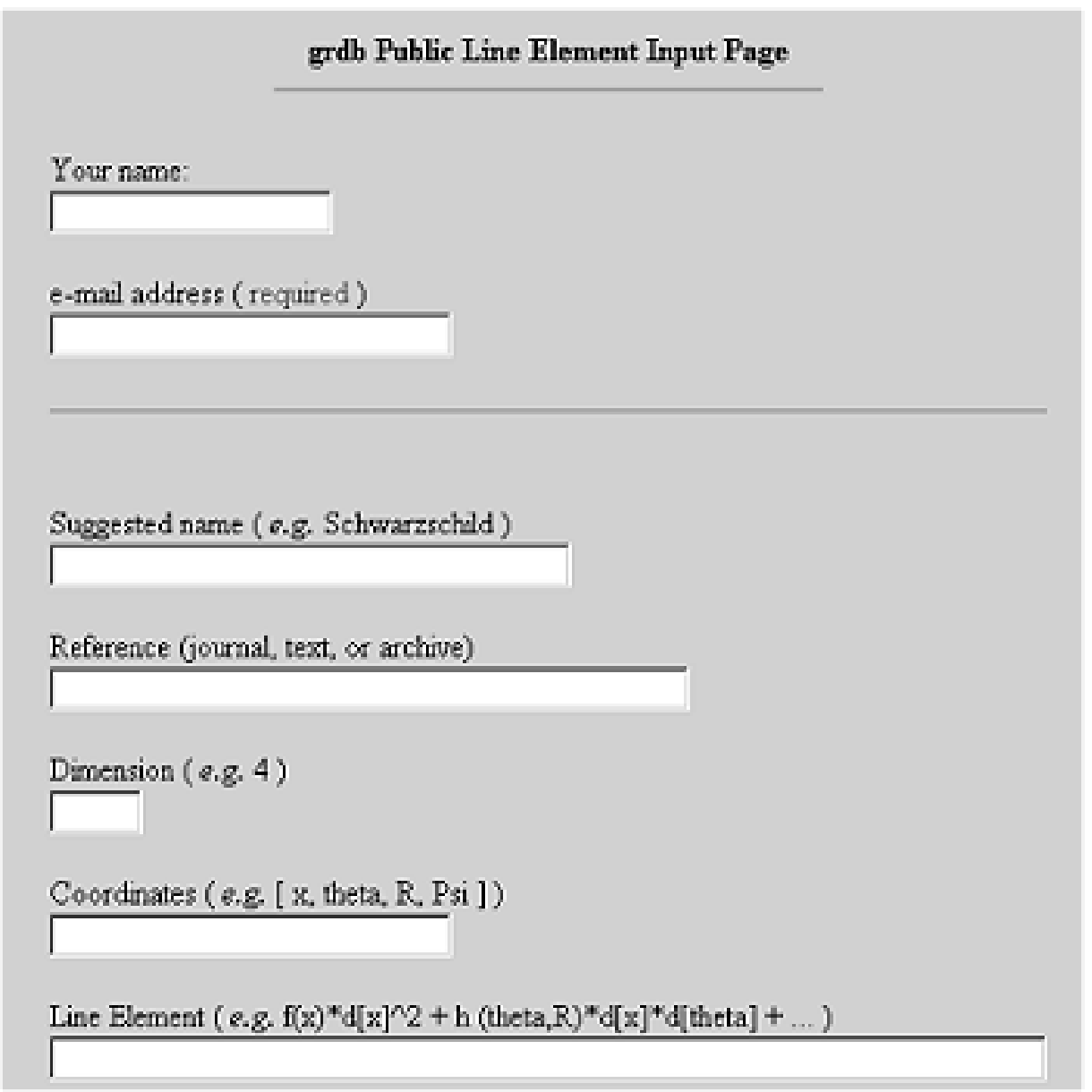}
\vspace{8pt} \caption{\label{publicinput} GRDB Public Line Element
Input Page. The fields are entered by the user following the
syntax provided in the examples shown at the top of each box.}

\end{figure}

\end{center}

GRDB offers many choices of operation. These operations are
typically carried in the following order:

\begin{itemize}

\item

i) Search the database by fields of particular interest. A search
by keywords or by an equation number e.g. as listed in the ``Exact
Solutions Book" \cite{kramer} is implemented. Other searches can
be easily added as the project evolves, e.g. a search by the
equation number as listed in any standard reference. The keyword
choice searches the information field and the name of the metric
field of each entry in the database. A common use is a search
based on an author's surname (e.g. Kruskal). The key word search
must be at least 3 characters. Reference searches follow a simple
referencing protocol. The search result is displayed in a list as
shown on the bottom of Figure \ref{homepage}. The record short
name is displayed as a hyper-link. The record naming convention is
explained in Appendix A.

\item
ii) Select a record or records of interest from the search result
list by clicking on its hyper-link. This will display a detailed
description of the space-time starting with a suggestive line
element representation of the record that is produced dynamically
(not stored). Record details are then displayed as in Figure
\ref{recordpage} with a structure adapted to the type of the
record examined (i.e. a metric, a tetrad or a basis type entry).

\item
iii) Mark the viewed record(s) to be loaded into the on line
calculator GRTensorJ. This calculator is described further bellow.

\item
iv) Click on the Start Calculator button from the database home
page in order to launch GRTensorJ and to do on line calculations
for the loaded records.

\item
v) Click on ``Load/Select'' from the menu inside the calculator
(Figure \ref{calculator}) and then click on ``Load Marked
Manifold(s) from Database''. The tagged manifolds will appear as
choices under Load/Select. Click on the manifold you wish to use
for calculations. This manifold is now loaded into the calculator
and the user can now perform calculations. An extensive list of
objects has already been implemented. This is summarized in
Appendix B. Customized calculations are described below.

\item

vi) Select the calculations commands through menu and sub-menu
selections in the calculator as shown in Figure \ref{calculator}.
The first step is to select a space-time in coordinates, tetrads
or basis. This choice automatically displays the components of the
metric tensor, the tetrads, or the basis components. Then, the
user can select the object(s) to be calculated from the
corresponding menus. After the result of the calculation is
displayed the user can apply to it any simplification procedure
supported by the engine as listed under ``Simplifications'' in the
menu bar.  A help system is built into the menu system. Further,
the user can define new objects through the definition facilities
within GRTensorII. These are entered through a sub-interface and
saved for the user's work session. There are also other support
and option selection functions. It is especially important to note
that if a manifold is represented with constraints (e.g Kruskal -
Szekeres) then the constraints (under ``Simplifications")
\textit{must} be applied on the calculations, otherwise the user
is not working with the selected manifold. A number of useful
features are easily added to GRTensorJ. For example, automatic
Latex has been implemented \cite{papers}.

\item
vii) New objects are defined in a sub-menu using the syntax of the
GRTensorII \texttt{grdef} facility \cite{grtensor}. For example,
the Bel-Robinson tensor (though pre-defined) would be entered as
$\grave{} \;  T\{(c d e f)\}:=C\{a c d b\}*C\{\hat{}\,a e
f\hat{}\,b\}+Cstar\{a c d b\}*Cstar\{\hat{}\,a e f \hat{}\,b\}\;
\grave{}$. In addition, there are objects that require an input
from the user, e.g. the kinematical quantities associated with a
field $u^{a}$. In order to create an object, select a manifold,
then select under the menu item ``Custom'' the option ``Create a
new single definition object''. Follow the syntax explained under
``Explanation'' and the example shown on the object input
interface, e.g. $\grave{} \;  u{^a}:=[u^1,u^2,u^3,u^4] \;
\grave{}$ where $u^1$ etc. are the components. Once the object is
created, it is saved during the user's work session and related
calculations can be performed, e.g. the acceleration, shear and
rotation for $u^{a}$. For a full description of the syntax the
user may consult GRTensorII \cite{grtensor} documentation on the
command ``grdef'' or browse the on line help system from the
GRTensorII WWW site \cite{grtensor}.

\item
viii) Input in the database. There are two forms of input into the
database, a Public Input and a Private Input. It is our intention
that all public input be refereed. Private input is password
protected and reserved for maintenance and internal operations.
The List-All option is currently for private usage. The public
input includes three types of interfaces; a line element, a
covariant or contravariant Newman-Penrose tetrad or a
non-holonomic basis. Figure \ref{publicinput} gives an example of
a line element input interface.

\end{itemize}

\section{Internal Design}

\subsection{GRDB components}

GRDB is built using fully modular Client/Server and
object-oriented designs. It uses Java Servlet technology
\cite{servlets}, a platform-independent mechanism for programming
applications over the WWW. Java is a pure object-oriented
language, allowing multi-thread processes, specially adapted to
network applications. Java Servlets are secure and portable
applications that handle requests from a Web server that are
originating from a client Web browser. The servlet runs on the
server side and does not depend on the browser compatibility. Java
Servlets also solve many access issue problems, as they can be
used to develop fully distributed applications over the Web where
the database can be on separate machines. GRDB also uses a JDBC
(Java DataBase Connectivity) module \cite{jdbc}, a portable and
flexible application that acts as an independent connector to the
database, increasing the modularity of the architecture.
PostgreSQL \cite{postgresql} was chosen and configured as the
DataBase Management System (DBMS ), a system that is also
object-oriented and that brings to GRDB all the power and
performance of the Structured Query Language (SQL) commands. This
language makes any kind of search or data manipulation very easy
to implement. The SQL instructions are formulated within the Java
Servlet code. The web-pages used in order to manipulate the Data
are dynamic pages created while the programs are running. Only the
help pages were left static. This HTML is also embedded in the
Servlet code. The records (manifolds) are for the most part stored
only in terms of a metric or set of basis vectors along with
constraints and referencing data. This information is processed
interactively through the Servlets and GRTensorJ which in turn
runs on Maple. The program TTH \cite{tth} is used to translate TEX
into HTML in order to produce mathematical html text.

\subsection{The Calculator GRTensorJ}

GRTensorJ is also written in Java and has been designed using a
full object-oriented approach. In addition, all communication
between the user and the server is object-based. GRTensorJ has a
multi-layer architecture that allows the generic functionality
described above. This architecture is outlined below:

\begin{itemize}

\item GUI (Graphical User Interface)
\item User Functional Interface
\item User ICM (Interchange Module) Handler
\item Interchange Modules
\item Server ICM Handler
\item Server Functional Interface
\item Algebraic Engine - Server Structure
\end{itemize}

A Java Applet (a programmed application that runs on the client
side) opens GRTensorJ by clicking on the applet button on GRDB
home page. It appears as in Figure \ref{calculator}. Behind the
scenes, a computer algebra session in GRTensorII is started
automatically. GRTensorJ uses Maple \cite{maple} as the algebraic
engine. GRTensorJ is designed with an open architecture and is
compatible with any other algebraic engine that can output an
ASCII stream. This architecture also allows it to be expanded and
programmed by the user. The commands are accessed through  menu
and sub-menu selections. When a session begins, GRTensorJ reads a
directory on the server named TextSheets (not to be confused with
``worksheets") and builds the menu and sub-menus for the interface
from the underlying structure. All sub-directories to TextSheets
will appear as primary menu bar items. The names of ASCII files
contained within these sub-directories will be displayed as menu
selections. These files contain a sequence of commands written in
the syntax of the computer algebra engine being used. By selecting
a menu item the user sends these commands to the engine. Items to
be displayed in the interface window are distinguished simply by
an asterisk in the file. In other words, creating new menu items
and calculation commands is as simple as creating and editing very
simple ASCII files. Yet, the file, and resultant menu item, can be
the equivalent of an entire worksheet with only the result chosen
for display. Of course, users need program nothing at all. The
underlying simplicity of GRTensorJ is mentioned here simply to
demonstrate its flexibility.

\section{Further Developments}

When first released (July 2001), the database contained only about
300 entries. This is under continuous development and, for
example, it is planed to have the entire content of \cite{kramer}
and \cite{krasin} included in the database. An extension of the
database to incorporate algorithmic classification is in progress
\cite{pollney} as is the construction of specialized record data
associated with, for example, static spherically symmetric perfect
fluid solutions \cite{delgaty}.

\section*{Acknowledgments}
This work was supported by a grant (to KL) from the Natural
Sciences and Engineering Research Council of Canada and by an
Ontario Graduate Scholarship (to MI).

\newpage

\begin{appendix}

\section{Record Naming Convention}

The record naming convention is not of significant interest to the
user as it is merely a placeholder for links. The convention does,
however, convey some information that can be quickly digested by
the user. The naming convention is as follows: ``abc''
\begin{itemize}
\item
1: No numbers
\item
2: All lower case
\item
3: a=name (e.g. bondi)
\item
4: b=m (metric), or npd (covariant tetrad), or npu (contravariant
tetrad), or b (basis)
\item
5: c=a,...,z (version)
\item
6: Default signature for metrics +2 for NP tetrads -2.
\end{itemize}
examples:
\\ \noindent kerrma means the Kerr spacetime in metric form version a
\\ \noindent kerrmb means the Kerr spacetime in metric form version b
\\ \noindent kerrnpda means the Kerr spacetime in NP tetrad
covariant form version a
\\ \noindent kerrnpua means the Kerr spacetime in NP tetrad
contravariant form version a.\\

\section{Initial set of preprogrammed objects available for calculation}
\newpage
\begin{table}
\begin{tabular}{lll}

\textbf{GRTensorJ menu item}  & \textbf{GRTensorJ sub-menu
command} & \textbf{Object(s)}\\ \\

Metric  & Metric                       & $g_{ab}$\\

        & Signature                    & $Sig$\\

        & Line Element                 & $ds^{2}$\\

        & Determinant of Metric        & $det(g)$\\

        & Inverse Metric               & $g^{ab}$\\

        & Partial Derivative of Metric & $g_{ab,c}$\\

Christoffel Symbols & Chr(dn,dn,dn)    & $\Gamma_{abc}$\\

                    & Chr(dn,dn,up)    & $\Gamma^{c}_{ab}$\\

Geodesic            &      &Writes out the geodesic equations      \\

Riemann             & R(dn,dn,dn,dn)   & $R_{abcd}$\\

                    & R(up,dn,dn,dn)   & $R^a\,_{bcd}$\\

                    & R(up,up,dn,dn)   & $R^{ab}\,_{cd}$\\

                    & R(up,up,up,up)   & $R^{abcd}$\\

                    & Kretschmann      & $R_{abcd}R^{abcd}$\\

Weyl                & C(dn,dn,dn,dn) ... & $C_{abcd}$\\

                    & Dual CStar(dn,dn,dn,dn)

... & $C^{*}_{abcd}\equiv\frac{1}{2}\epsilon_{abef}C^{ef}\,_{cd}$ \\

Ricci               & R(dn,dn) ...     & $R_{ab}\equiv R^c\,_{acb}$\\

Trace-free Ricci    & S(dn,dn) ...     & $S_{ab}\equiv R_{ab}-\frac{1}{N}g_{ab}R$\\

Einstein            & G(dn,dn) ...     & $G_{ab}\equiv R_{ab}-\frac{1}{2}g_{ab}R$\\

Invariants          & Invariants-Ricci & $R\equiv R^{a}\,_{a},\, R1\equiv\frac{1}{4}S^{a}\,_{b}S^{b}\,_{a}, \; R2\equiv\frac{-1}{8}S^{a}\,_{b}S^{b}\,_{c}S^{c}\,_{a}$ \\

                    &                  & $R3\equiv\frac{1}{16}S^{a}\,_{b}S^{b}\,_{c}S^{c}\,_{d}S^{d}\,_{a}$ \\

                    & Invariants-Weyl  & $W1R\equiv\frac{1}{8}C_{abcd}C^{abcd}, \; W1I\equiv\frac{1}{8}C^{*}\,_{abcd}C^{abcd}$ \\

                    &                  & $W2R\equiv\frac{-1}{16}C_{ab}\,^{cd}C_{cd}\,^{ef}C_{ef}\,^{ab}$ \\

                    &                  & $W2I\equiv\frac{-1}{16}C^{*}\,_{ab}\,^{cd}C_{cd}\,^{ef}C_{ef}\,^{ab}$ \\

                & Invariants-Mixed & $M1R \equiv \frac{1}{8}S^{ab}S^{cd}C_{abcd},\, M1I\equiv

\frac{1}{8}S^{ab}S^{cd}C^{*}\,_{abcd}$ \\

                    &                  & $M2R \equiv \frac{1}{16}S^{cd}S_{ef}(C_{acdb}C^{aefb}-C^{*}_{acdb}C_{*}^{aefb})$ \\

                    &                  & $M2I \equiv \frac{1}{8}S^{bc}S_{ef}(C^{*}_{abcd}C^{aefd})$ \\

                    &                  & $M3 \equiv \frac{1}{16}S^{cd}S_{ef}(C_{acdb}C^{aefb}+C^{*}_{acdb}C_{*}^{aefb})$ \\

                    &                  & $M4 \equiv \frac{-1}{32}S^{cg}S^{ef}S^{c}\,_{d}(C_{ac}\,^{db}C_{befg}+C^{*}_{ac}\,^{db}C^{*}_{befg})$ \\

              &                  & $M5R \equiv \frac{1}{32}S^{cd}S^{ef}C^{aghb}(C_{acdb}C_{gefh}+C^{*}_{acdb}C^{*}_{gefh})$\\

                    &                  & $M5I \equiv \frac{1}{32}S^{cd}S^{ef}C_{*}\,^{aghb}(C_{acdb}C_{gefh}+C^{*}_{acdb}C^{*}_{gefh})$\\

                    &                  & $M6R \equiv \frac{1}{32} S_a{}^e S_e{}^c S_b{}^f

                  S_f{}^d C^{ab}{}_{cd}$  \\

                    &                 & $ M6I \equiv \frac{1}{32} S_a{}^e S_e{}^c S_b{}^f

                  S_f{}^d C^{*ab}{}_{cd}$  \\

Differential Invariants& diRicci         & $R_{ab;c}R^{ab;c}$\\

                       & diRiem          & $R_{abcd;e}R^{abcd;e}$\\

                       & diS             & $S_{ab;c}S^{ab;c}$\\

                       & diWeyl          & $C_{abcd;e}C^{abcd;e}$\\

Bel-Robinson           & T(dn,dn,dn,dn) ... &

$T_{cdef}\equiv C_{acdb}C^{a}\,_{ef}\,^{b}+C^{*}\,_{acdb}C^{*}\,^{a}\,_{ef}\,^{b}$ \\

Weyl-Schouten          & Weyl-Schouten ...  &

$WS_{abc}\equiv R_{ab;c}-R_{ac;b}-( g_{ab}R^{e}\,_{e;c}-g_{ac}R^{f}\,_{f;b} )/4$ \\

Bach                   & B(dn,dn) ...       & $B_{ac}\equiv

C_{abcd}\,^{;bd}+\frac{1}{2}R^{bd}C_{abcd}$\\

Input field calculations& Field u         & $u^{a}$\\

                        & KillingCoords   & Adapted coordinate check    \\

                        & KillingTest     &  Test for conformal/homothetic/Killing vector  \\

                        & Vector Kinematics & $\dot{u}^a\equiv

                        u^{b}u^{a}\,_{;b}, \; \theta\equiv u^{b}\,_{;b}, \;

                        \sigma_{a}\,^{b}\equiv

                        u_{(a}\,^{;b}\,_{)}+\dot{u}_{(a}u^{b}\,_{)}-$ \\

                        &                 &$\frac{1}{3}h_{a}\,^{b}\dot{\theta},\,\,\,\omega_{ab}\equiv u_{[a;b]}+\dot{u}_{[a}u_{b]}$\\

                        & Electric Weyl E(dn,dn) ...& $E_{ac}\equiv

C_{abcd}u^{b} u^{d}$ \\

                        & Magnetic Weyl H(dn,dn) ...& $H_{ac}\equiv

C^{*}\,_{abcd}u^{b} u^{d}$\\

\\

\end{tabular}
\caption{Initial set of preprogrammed objects available for
coordinate calculations. Note: "..." means that all index
combinations are immediately available as with Riemann.}
\end{table}

\begin{table}

\begin{tabular}{lll}

\textbf{GRTensorJ menu item } & \textbf{GRTensorJ sub-menu
command} & \textbf{Object(s)} \\ \\

NPSpin &  NPSpin: NPkappa, NPsigma, NPlambda, & spin coefficients, \\

       &  NPnu, NPrho, NPmu, NPtau, NPpi,     & $\kappa, \sigma, \ldots, \beta$\\

       &  NPepsilon, NPgamma, NPalpha, NPbeta & \\

\\

NPSpinbar & NPSpinbar: NPkappabar, NPsigmabar, & spin coefficients\\

          & NPlambdabar, NPnubar, NPrhobar,    & (complex conjugates) \\

          & NPmubar, NPtaubar, NPpibar, NPepsilonbar, & $\bar{\kappa}, \bar{\sigma},

                                 \ldots \bar{\beta}$ \\

          & NPgammabar, NPalphabar, NPbetabar &  \\

\\

Ricci Scalars & Ricci Scalars: Phi00, Phi01, Phi02, & Ricci scalars, \\

              & Phi10, Phi11, Phi12, Phi20, Phi21,  & $\Phi_{00}, \Phi_{01}, \ldots,

                                                      \Phi_{22},\Lambda$ \\

              & Phi22, Lambda                       & \\

\\

Weyl Scalars & Weyl Scalars: Psi0, Psi1, Psi2, Psi3, Psi4 & Weyl
scalars,

                                           $\Psi_0, \ldots, \Psi_4$\\

\\

             & Petrov                           & Petrov type \\

             & Petrov Report                    & A report on how the\\

             &                                      & Petrov type was determined \\

\\

\end{tabular}

\caption{Initial set of preprogrammed objects available for tetrad
calculations. All the objects in TABLE I are integrated with
tetrad input and can be run directly from a tetrad.}

\end{table}

\begin{table}

\begin{tabular}{lll}

\textbf{GRTensorJ menu item}  & \textbf{GRTensorJ sub-menu
command} & \textbf{Object(s)} \\ \\

Rotation Coefficients & lambda(bdn,bdn,bdn) &
$\lambda_{(a)(b)(c)}:=

                             e_{(b)[i,j]} e_{(a)}{}^d e_{(c)}{}^e$ \\

                      & rot(bdn,bdn,bdn)    & rotation coefficients, \\

                      &                     & $\gamma_{(a)(b)(c)} := \frac{1}{2} \left(

                                \lambda_{(a)(b)(c)}+\lambda_{(c)(a)(b)} -

                                \lambda_{(b)(c)(a)} \right)$ \\

                      & str(bdn,bdn,bdn)    & structure constants, \\

              &                     & $C_{(a)(b)(c)} := \gamma_{(a)(c)(b)}

                                  - \gamma_{(a)(b)(c)}$ \\

\\

\end{tabular}

\caption{Initial set of preprogrammed objects available for basis
calculations. All the objects in TABLE I are integrated with basis
input and can be run directly from a basis.}

\end{table}
\end{appendix}

\newpage

\begin{center}
\begin{figure}
\includegraphics{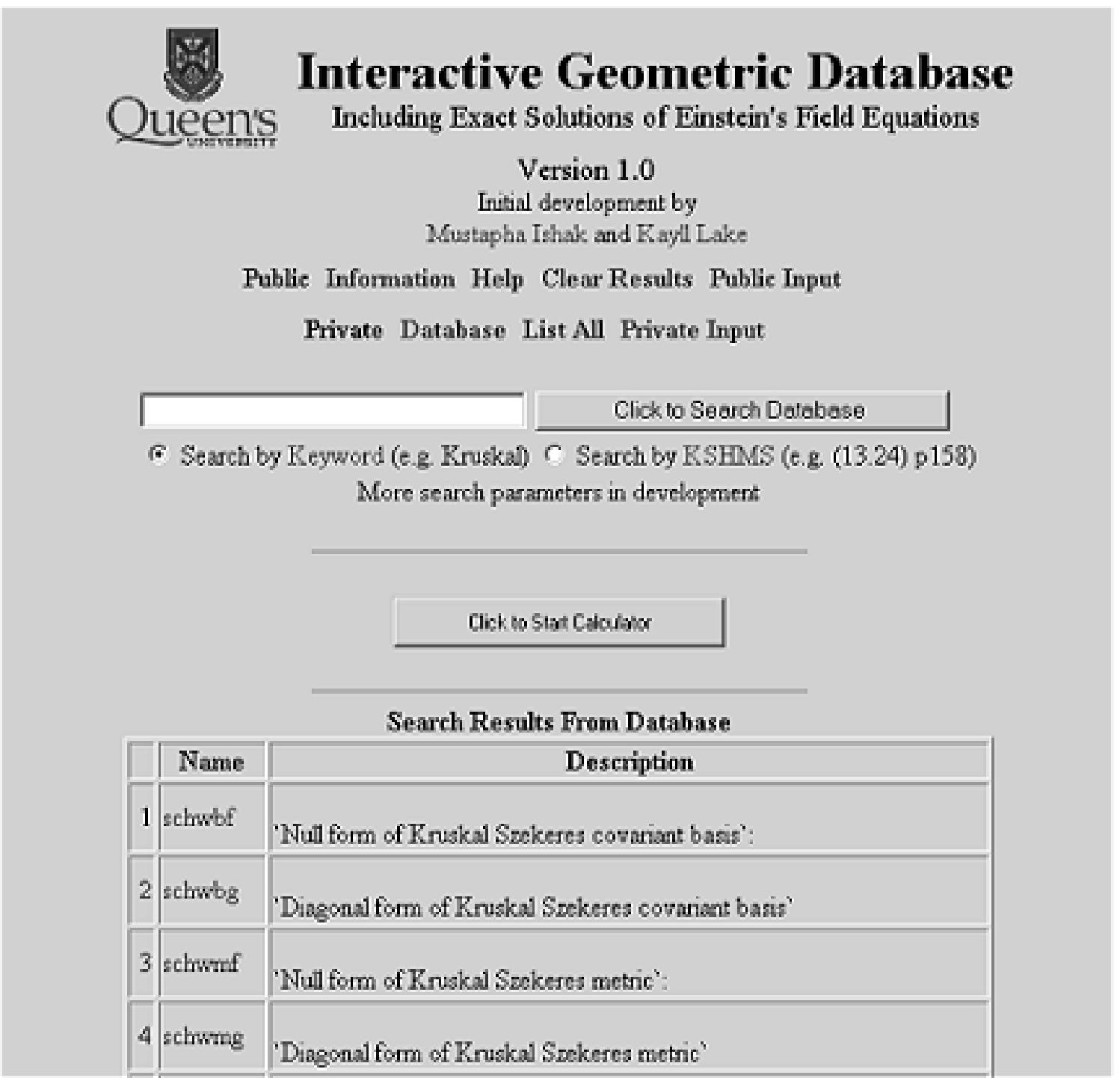}
\caption{\label{homepage} GRDB Home Page (at the WWW address
http://grdb.org). The operations on the database available at the
time this paper was written are displayed, followed by the search
form and the Calculator Applet button. Part of the search result
by the key word ``Kruskal'' is displayed.}
\end{figure}
\end{center}

\newpage

\newpage

\newpage

\begin{thebibliography}{}
\bibitem{papers}A latex call in Maple produces output suitable for use with LaTeX 2e and appears in the applet window. We are
serious when we say that you can, for example, now go to the Web,
do your calculation, and paste the answer into the paper you are
writing!
\end{thebibliography}
\end{document}